\documentclass[pdflatex,sn-mathphys]{sn-jnl}

\jyear{2021}

\begin{document}

\title[]{Environmental Memory Boosts Group Formation of Clueless Individuals}

\author[1,2]{\fnm{Cristóvão S.} \sur{Dias}}

\author[3]{\fnm{Manish} \sur{Trivedi}}

\author*[4]{\fnm{Giovanni} \sur{Volpe}}\email{giovanni.volpe@physics.gu.se}

\author*[1,2]{\fnm{Nuno A. M.} \sur{Araújo}} \email{nmaraujo@fc.ul.pt}

\author*[3]{\fnm{Giorgio} \sur{Volpe}}\email{g.volpe@ucl.ac.uk}

\affil[1]{\orgdiv{Departamento de Física, Faculdade de Ciências}, \orgname{Universidade de Lisboa}, \orgaddress{\postcode{1749-016} \city{Lisboa}, \country{Portugal}}}

\affil[2]{\orgdiv{Centro de Física Teórica e Computacional, Faculdade de Ciências}, \orgname{Universidade de Lisboa}, \orgaddress{\postcode{1749-016} \city{Lisboa}, \country{Portugal}}}

\affil[3]{\orgdiv{Department of Chemistry}, \orgname{University College London}, \orgaddress{\street{20 Gordon Street}, \postcode{WC1H 0AJ} \city{London}, \country{United Kingdom}}}

\affil[4]{\orgdiv{Department of Physics}, \orgname{University of Gothenburg}, \orgaddress{\street{Origovägen 6B}, \postcode{SE-41296}  \city{Gothenburg}, \country{Sweden}}}

\abstract{The formation of groups of interacting individuals improves performance and fitness in many decentralised systems, from micro-organisms to social insects, from robotic swarms to artificial intelligence algorithms. Often, group formation and high-level coordination in these systems emerge from individuals with limited information-processing capabilities implementing low-level rules of communication to signal to each other. Here, we show that, even in a community of clueless individuals incapable of processing information and communicating, a dynamic environment can coordinate group formation by transiently storing memory of the earlier passage of individuals. Our results identify a new mechanism of indirect coordination via shared memory that is primarily promoted and reinforced by dynamic environmental factors, thus overshadowing the need for any form of explicit signalling between individuals. We expect this pathway to group formation to be relevant for understanding and controlling self-organisation and collective decision making in both living and artificial active matter in real-life environments.} 

\maketitle

\clearpage

``Strength in numbers'' is more than an idiomatic expression. Many living systems form groups to improve their fitness, optimise use and allocation of resources, and reach consensus \cite{vicsek2012collective}. Examples emerge at all length scales, from 
bacterial quorum sensing and biofilm formation \cite{mukherjee2019bacterial} to social insects \cite{czaczkes2015trail}, from animal groups \cite{moussaid2009collective} to human crowds \cite{sieben2017collective}. Artificial active matter systems, such as active colloids \cite{bechinger2016active} and robotic swarms \cite{dorigo2020reflections}, provide controllable systems to pinpoint the essential principles behind the emergence of these collective behaviours in living systems \cite{Palacci2012,Lavergne2019}. For example, active colloids have been employed to demonstrated motility-induced phase separation \cite{Buttinoni2013,Cates2015} as well as the spontaneous formation of living crystals resembling animal group formation \cite{Palacci2012,Ginot2018}. Complex dynamic collective patterns, such as colloidal swarms, flocks and swirls, have also been demonstrated by introducing controllable attractive, repulsive or aligning interactions among individuals by particle design \cite{Bricard2013,Granick2016}, by defining appropriate confining potentials \cite{Pince2016} or by modulating particles' propulsion with external feedback loops \cite{khadka2018,Lavergne2019}. In recent years, a few active particles in crowded environments of passive colloids have also been employed to modulate the energy landscape of the passive phase with an emphasis on controlling the assembly of soft materials \cite{reichhardt2004local,kummel2015formation,van2016fabricating,omar2018swimming,dietrich2018active,ramananarivo2019activity,banerjee2022unjamming,madden2022hydrodynamically,trivedi2022self}.

Whether living or artificial, decentralised systems are characterised by high-level coordination and collective behaviours, which emerge from individuals with limited information-processing capabilities responding to low-level rules of engagement \cite{garnier2007biological}. In particular, stigmergy is a form of indirect communication between individuals mediated by modifications of the local environment, where individuals actively signal to others by depositing cues which shape a shared environmental memory \cite{marsh2008stigmergic}. This strategy underpins the emergence of coordination and collective decision making in many natural decentralised systems, from micro-organisms \cite{mukherjee2019bacterial} to social insects \cite{czaczkes2015trail}. For example, trailing stalk cells guided by chemo-attractants through tissue establish the vascular lumen in sprout angiogenesis \cite{Lugano2020}; bacteria \cite{Austin2020}, amoebas \cite{Tweedy2020} and ants \cite{Reid2011} can solve physical mazes by tracking chemical scents and forming optimal paths; mutual anticipation and avoidance in crowds lead to lane formation and stabilisation \cite{murakami2021mutual}. The concept of stigmergy has also found widespread use in technological and engineering applications, from robotic swarms \cite{werfel2014designing} to artificial intelligence algorithms \cite{dorigo2020reflections}, to, recently, active colloids \cite{nakayama2023tunable}. In these systems, it is usually assumed that individuals possess a minimal level of low-level communication and signal processing capabilities, which leads to the emergence of a shared environmental memory and, eventually, high-level group dynamics \cite{marsh2008stigmergic}.

Here, we demonstrate that, even in a community of clueless self-motile individuals (i.e., incapable of directly signalling to each other or processing information), avoidance of a dense population of non-fixed obstacles is sufficient to lead to the emergence of stigmergy when the dynamic environment can transiently store memory of the earlier passage of individuals. Counterintuitively, we find that, while the motion of the individuals is hampered by increasing levels of crowding, the spatial correlations created and stored in the otherwise passive environment after their passage feed back on the motion of other individuals boosting aggregation rates and, consequently, group formation.

\begin{figure}[!]%
\centering
\includegraphics[width=0.5\textwidth]{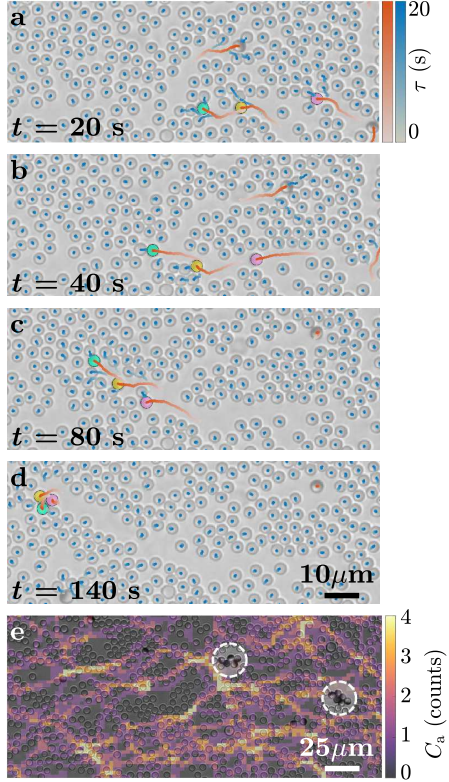}
\caption{\textbf{Group formation of active colloids mediated by environmental memory.}  (\textbf{a-d}) Time sequence showing (\textbf{a}) a light-activated Janus particle (cyan) forming a path in a crowded environment of ${\rm SiO_2}$ passive particles (densities: $\rho_{\rm a} = 1.1\%$ for active particles and $\rho_{\rm p} = 37.5\%$ for passive particles), (\textbf{b-c}) which is then reused by nearby Janus particles (yellow and magenta) leading to (\textbf{d}) the formation of a group (here, a three-particle cluster). In each image, 20-s-long trajectories are shown for both active (red) and passive (blue) particles. (\textbf{e}) Counts $C_{\rm a}$ (represented as a heatmap) of individual Janus particles that have transited on a pixel during a 16-minute acquisition in a similar crowded environment as in \textbf{a-d}. The heatmap is overlaid to the final frame showing how path generation and reuse (bright lines) correlate to group formation and cohesion of active units in time (white dashed circles). The heatmap was obtained from a sample with $\rho_{\rm a} = 1.1\%$ and $\rho_{\rm p} = 37.5\%$ and an occupied pixel was only accounted for once for each particle.}
\label{fig1}
\end{figure}

As paradigmatic self-motile individuals, we employ Janus silica (${\rm SiO}_2$) colloids (diameter $d = 4.77 \pm 0.20 \, {\rm \mu m}$) half-coated with a thin layer of carbon ($\approx 60 \, {\rm nm}$) (Methods). When suspended in a critical binary mixture of water and 2,6-lutidine ($0.286$ mass fraction of lutidine) below its lower critical temperature ($T_{\rm c} \approx 307 \, {\rm K}$), these colloids undergo Brownian diffusion \cite{buttinoni2012active}. Upon exposure to laser illumination ($\lambda = 532 \, {\rm nm}$, $I \approx 2.5 \, {\rm \mu W \mu m^{-2}}$) (Methods), light absorption at the carbon cap drives the Janus particles' self-diffusiophoretic motion at a speed of $v \approx 1.9 \, {\rm \mu m \, s^{-1}}$ due to local heating and demixing of the critical mixture around the cap \cite{buttinoni2012active}. Because of their colloidal nature, these self-motile individuals are clueless in the sense that they have no sensing or information processing capabilities and interact with each other through simple physical interactions rules, such as steric and short-range attractive interactions  \cite{Palacci2012,Buttinoni2013,mognetti2013living}. Boundaries can also influence their motion with aligning interactions  \cite{das2015boundaries,simmchen2016topographical}.  

To study their interplay with a dynamic environment of non-static obstacles, we prepare quasi-two-dimensional samples of Janus particles mixed with dispersions of equally-sized freely diffusing silica (${\rm SiO}_2$) colloids at different densities ($0 \le \rho_{\rm p} \le 75 \%$, defined as fractional surface coverage), where the Janus particles only represent a small portion ($0.5 \% \le \rho_{\rm a} \le 1.6 \%$, also defined as fractional surface coverage) (Methods).  The two example time sequences of active particles ($\rho_{\rm a} = 1.1\%$) moving in a crowded environment ($\rho_{\rm p} = 37.5\%$) in Figs. \ref{fig1} and \ref{fig:Extra_Fig1_time_sequence} show how the changes introduced in the passive phase by the active colloids produce spatial correlations in the environment in the form of open transient paths. These paths feed back on the motion of the active particles, eventually leading to group formation (here defined as the formation of a stable cluster of at least three particles). While moving forward, individual Janus particles need to physically dig their own path against the surrounding background of passive colloids (Figs. \ref{fig1}a and \ref{fig:Extra_Fig1_time_sequence}a), which in turn reduces their overall motility for increasing values of $\rho_{\rm p}$ (as exemplified by the mean square displacements in Fig. \ref{fig:MSD}). Interestingly, before closing due to the Brownian motion of the passive colloids, these paths appear to be reused by nearby active colloids, which favour reusing these preformed paths of lower resistance rather than digging their owns (Figs. \ref{fig1}b and \ref{fig:Extra_Fig1_time_sequence}b). A form of stigmergy between the active particles is then established thanks to  their passive counterparts, where the transient paths opened by the active colloids in their surroundings become a shared environmental memory for their peers, which generates a feedback that reinforces their trailing behaviour and, eventually, leads to group formation. Indeed, over time, the trailing Janus colloids catch up with the front particles (Figs. \ref{fig1}c and \ref{fig:Extra_Fig1_time_sequence}c) to form a small cluster (Figs. \ref{fig1}d and \ref{fig:Extra_Fig1_time_sequence}d). Once formed, these clusters then grow to larger sizes due to the continuous addition of new active particles to the group through a network of similar transient open paths that form and evolve over time (Fig. \ref{fig1}e).

\begin{figure}[h!]%
\centering
\includegraphics[width=1\textwidth]{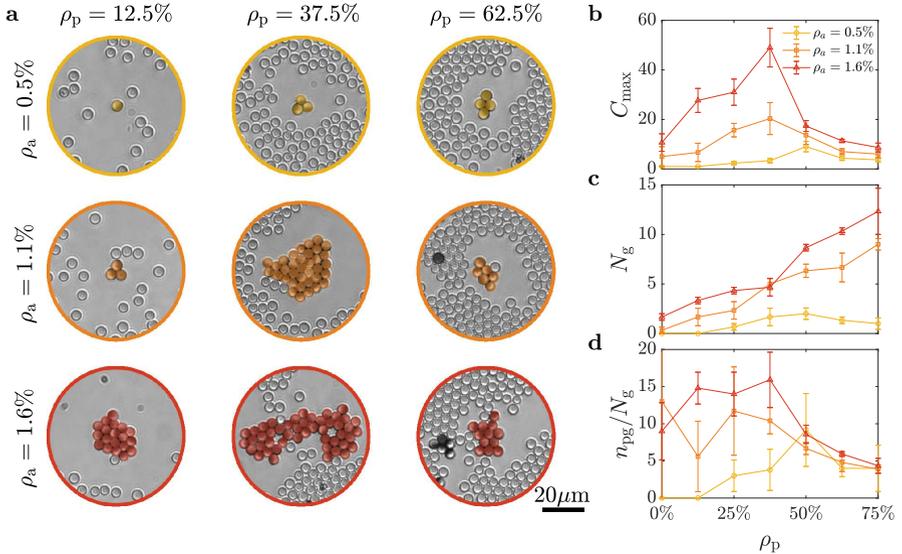}
\caption{\textbf{Non-monotonic size dependence of group formation on environmental crowding.} Groups (here defined as all stable clusters formed by at least 3 individuals) formed after 25 minutes as a function of the initial densities of active ($\rho_{\rm a}$) and passive ($\rho_{\rm p}$) particles. (\textbf{a}) Example images of the largest clusters of active colloids formed at different $\rho_{\rm a}$ and $\rho_{\rm p}$. 
Independent of $\rho_{\rm a}$, these images show how the largest sizes are obtained at intermediate values of $\rho_{\rm p}$. This visual trend is confirmed by (\textbf{b}) the non-monotonic dependence of the average size $C_{\rm max}$ of the largest clusters (measured as number of active particles) as a function of $\rho_{\rm p}$ for different values of $\rho_{\rm a}$. (\textbf{c}) Differently from $C_{\rm max}$, the total number of groups $N_{\rm g}$ tends to monotonically increase with $\rho_{\rm p}$. The error bars in \textbf{b} and \textbf{c} represent one standard error around the average values obtained from triplicates. (\textbf{d}) Average number of active colloids in a group after 25 minutes calculated as the ratio between number of particles in a group $n_{\rm pg}$ and the number of groups $N_{\rm g}$. The error bars in \textbf{d} are obtained by propagating the standard errors on $n_{\rm pg}$ and $N_{\rm g}$.   
}\label{fig2}
\end{figure}

Figure \ref{fig2}a shows examples of the largest groups obtained after 25 minutes of experimental time for different values of $\rho_{\rm a}$ and $\rho_{\rm p}$. Independently of $\rho_{\rm a}$, the largest groups appear to form for intermediate values of $\rho_{\rm p}$, where the crowding is sufficient to create a shared memory in the environment in the form of reusable transient paths, but not dense enough to completely hamper the motility of the individual Janus particles. To quantify this observation, we calculated the size of the largest cluster $C_{\max}$ (Fig. \ref{fig2}b), the total number of groups $N_{\rm g}$ 
(Fig. \ref{fig2}c) and the average number of particles per group (Fig. \ref{fig2}d) for different values of  $\rho_{\rm a}$ as a function of $\rho_{\rm p}$. At $\rho_{\rm p} = 0$, no group forms at low density of active particles ($\rho_{\rm a} = 0.5\%$) as encounters are sparse, while the formation of a very few groups (up to $\approx 2$) becomes increasingly more likely for higher values of $\rho_{\rm a}$, as chances for encounter increase with the number of available individuals. Increasing $\rho_{\rm p}$ to intermediate values (up to $37.5\%$) leads to the formation of more groups on average (up to $\approx 5$, Fig. \ref{fig2}c). While, for a given $\rho_{\rm a}$, the average size of these groups seems roughly constant (Fig. \ref{fig2}d), a larger cluster emerges (Fig. \ref{fig2}b) that can contain up to $\approx 67\%$ of the particles in a group due to the shared environmental memory from the path reuse highlighted in Figs. \ref{fig1} and \ref{fig:Extra_Fig1_time_sequence}. This reuse can be quantified through the path revival function $1-C_{\rm aa}(\tau)$, where $C_{\rm aa}(\tau)$ is the cumulative probability that a region crossed by an active particle will be crossed by another particle within a lag time $\tau$ (Methods). The faster $1-C_{\rm aa}(\tau)$ decays (i.e., the faster $C_{\rm aa}(\tau)$ increases to one), the sooner a region explored by a particle will be crossed by another particle. Figure~\ref{fig:lifetime} shows how, in our experiments, the path revival function at first decays faster at higher $\rho_{\rm p}$ (quantified by the path revival lifetime $\tau_{\rho_{\rm p}}$), thus indicating a higher likelihood of reusing previously explored region. This trend of the path revival lifetime with $\rho_{\rm p}$ is unexpected and can only by justified by the emergence of a shared environmental memory due to the reuse of pre-existing paths. In fact, if we consider collisions between persistent particles whose velocities are Poisson distributed \cite{enns1984hitting}, we would expect the lifetime of the path revival function $\tau_{\rho_{\rm p}}$ to increase with $\rho_{\rm p}$ as the particles' effective velocity decreases due to the collisions with the passive particles (as confirmed by the MSDs in Fig. \ref{fig:MSD}) --- a trend that in our data only appears for high densities ($\rho_{\rm p} > 50\%$, Fig. \ref{fig:lifetime}). Indeed, an even further increase of $\rho_{\rm p}$ has a dramatic effect on group formation, as the reduced motility for the active colloids due to the resistance offered by the passive particles (Fig. \ref{fig:MSD}) induces a more intuitive behaviour where group formation and cohesion are drastically hampered by the crowded environment and any reduction of the path revival lifetime with respect to the case at $\rho_{\rm p} = 0 \%$ (Fig. ~\ref{fig:lifetime}) is now driven by the active particles being more localised in space due to crowding rather than the presence of longer-range correlations in the form of transient paths. For a given $\rho_{\rm a}$, the ever-increasing number of groups $N_{\rm g}$ with $\rho_{\rm p}$ (Fig. \ref{fig2}c) translates now into smaller groups of more homogeneous sizes and more localised in space, which, differently from $N_{\rm g}$, are roughly independent of the initial value of $\rho_{\rm a}$ (Figs. \ref{fig2}b and \ref{fig2}d).

\begin{figure}[htbp]%
\centering
\includegraphics[width=0.8\textwidth]{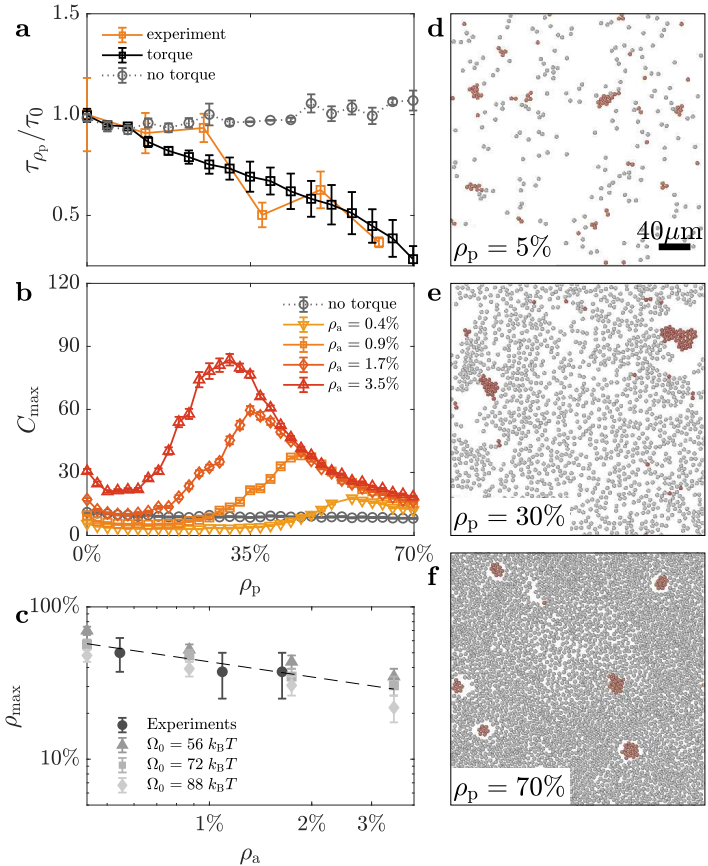}
\caption{\textbf{Importance of aligning interactions for stigmergy of active particles.} (\textbf{a}) Path revival time $\tau_{{\rho}_{\rm p}}$ as a function of $\rho_{\rm p}$ in experiments for $\rho_{\rm a} = 1.1\%$ (orange solid line) and simulations with (black solid line) and without (gray dotted line) an effective torque steering the active particles away from the passive ones. The torque ($\Omega_0 = 72 \pm 16 \, k_{\rm B}T$) is necessary to reproduce the experimental trend. The data from Fig. \ref{fig:lifetime} are here normalised to $\tau_{0}$, the survival time at $\rho_{\rm p} = 0$. (\textbf{b}) Simulated average size $C_{\rm max}$ of the largest group 
for different $\rho_{\rm a}$ as a function of $\rho_{\rm p}$ in the presence (black solid line) and absence (gray dotted line) of a torque, showing that the torque is key to the appearance of a peak at intermediate values of $\rho_{\rm p}$ (as in Fig. \ref{fig2}b). (\textbf{c}) The value of $\rho_{\rm p}$ at the peak position ($\rho_{\rm max}$) as a function of $\rho_{\rm a}$ provides a second estimate for the experimental torque strength ($\Omega_0 = 72 \pm 16 \, k_{\rm B}T$). Its monotonously decreasing trend is visualised by a dashed line as a guide for the eyes. Each experimental data point in {\bf a} and {\bf c} is obtained as an average from three videos at the corresponding values of $\rho_{\rm a}$ and $\rho_{\rm p}$ and simulations in {\bf a-c} are averages over 100 numerical experiments per value of $\rho_{\rm a}$ and $\rho_{\rm p}$. All error bars represent one standard error around the average values. (\textbf{d-f}) Example snapshots from simulations showing group formation of active particles (red, $\rho_{\rm a} = 1.1\%$) at different densities of passive particles: (\textbf{d}) $\rho_{\rm p} = 5\%$, (\textbf{e}) $\rho_{\rm p} = 30\%$ and (\textbf{f}) $\rho_{\rm p} = 70\%$.
}
\label{fig3}
\end{figure}

To gain a microscopic understanding of the non-monotonic dynamics of group formation in Fig. \ref{fig2}b, we consider a simple particle-based model that includes an aligning torque in the equations of motion of the active particles (Methods) \cite{Liebchen2019}. The effect of this torque is to steer the active particles away from the surrounding passive ones, and align their direction of motion to any effective boundary of a transient path \cite{das2015boundaries,simmchen2016topographical}. Figure \ref{fig3}a shows how the presence of this aligning torque is fundamental to reproduce the dependence of the experimental path revival time $\tau_{\rho_{\rm p}}$ on $\rho_{\rm p}$, indicating that obstacle avoidance is the mechanism that promotes active particles
to follow previously formed transient paths. In fact, in the absence of the torque, the revival lifetime increases with $\rho_{\rm p}$ as one would intuitively expect due to the decrease of the particles' effective velocity caused by collisions with the obstacles \cite{enns1984hitting}. Fitting our experimental data to our model allows us to determine the strength of the torque to be $\Omega_0 = 72 \, \pm 16 \,  k_{\rm B}T$. As shown in Fig. \ref{fig3}b, the presence of this torque is also critical to recover the non-monotic dependence of the largest group size with $\rho_{\rm p}$ as observed in our experimental data (Fig. \ref{fig2}b). Therefore, this torque and the resulting aligning interaction provide an enabling mechanism for the emergence of stigmergy via a shared environmental memory in the system of non-communicating active particles by allowing the spatial correlations in the environment to feed back on the active particles' motion. Figure \ref{fig3}b also confirms that the non-monotonic dynamics of group formation depend on the numbers of individuals (as already observed in Fig. \ref{fig2}b). At higher values of $\rho_{\rm a}$, encounters become more probable so that groups can form and grow to larger sizes at lower values of $\rho_{\rm p}$. As groups grow to larger sizes, the density of passive particles needed to cage them and prevent them from merging into even larger groups lowers too. The combination of these two effects translate in a decreasing monotonic dependence of the peak position on $\rho_{\rm a}$ (Fig. \ref{fig3}c), highlighting the relevance of environmental memory effects for group formation in sparse systems of clueless active particles. Figures \ref{fig3}d-f show example snapshots from the simulations, which confirm our qualitative observations in Fig. \ref{fig2}: at low values of $\rho_{\rm p}$ (Fig. \ref{fig3}d), groups of a few units are formed; at intermediate values of $\rho_{\rm p}$ (Fig. \ref{fig3}e), the landscape is dominated by a very few large groups that collect most of the active particles; finally, at large values of $\rho_{\rm p}$ (Fig. \ref{fig3}f), a few relatively smaller groups of more homogeneous size appear to be caged within the crowded environment. 

\begin{figure}[htbp]%
\centering
\includegraphics[width=0.5\textwidth]{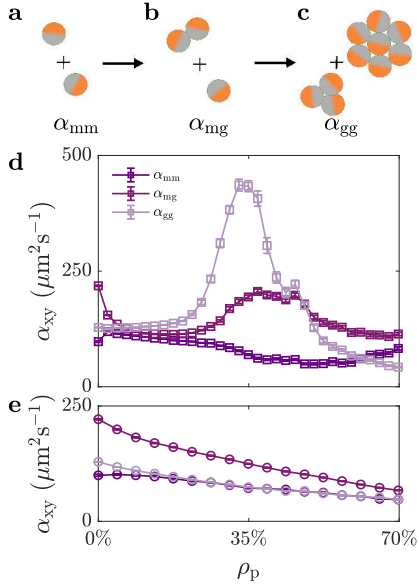}
\caption{\textbf{Kinetics of group formation with shared environmental memory.} (\textbf{a-c}) Schematics of the three main mechanisms for the kinetics of group formation described by the relevant rate coefficients $\alpha_{\rm xy}$: (\textbf{a}) monomer--monomer ($\alpha_{\rm mm}$), (\textbf{b}) monomer--group ($\alpha_{\rm mg}$) and (\textbf{c}) group--group ($\alpha_{\rm gg}$) aggregation.
(\textbf{d-e}) Calculated monomer--monomer ($\alpha_{\rm mm}$), monomer--group ($\alpha_{\rm mg}$) and group--group ($\alpha_{\rm gg}$) aggregation rate coefficients at $\rho_{\rm a} = 1.1\%$ as a function of $\rho_{\rm p}$ (\textbf{d}) in the presence of a torque and (\textbf{e}) in its absence. Simulations were averaged over 100 numerical experiments per value of $\rho_{\rm p}$. Error bars represent one standard error around the average values.}\label{fig4}
\end{figure}

To further understand how the shared environmental memory affects encounter dynamics, we can define a kinetic model based on mean-field rate equations for the number density of monomers (free active particles, i.e., not part of a group) $c_1$ and for the number density of groups $c_\mathrm{g}$ as
\begin{equation} \label{eq:dynamics_monomer} 
\dot{c}_1=-\alpha_\mathrm{mm}c^2_1-\alpha_\mathrm{mg}c_1c_{\rm g}, \\
\end{equation}
\begin{equation}\label{eq:dynamics_group} 
\dot{c}_{\rm g}=\frac{\alpha_\mathrm{mm}}{2}c_1^2-\frac{\alpha_\mathrm{gg}}{2}c_{\rm g}^2,
\end{equation}
where $\alpha_\mathrm{mm}$, $\alpha_\mathrm{mg}$ and $\alpha_\mathrm{gg}$ are the rate coefficients of monomer-monomer, monomer-group and group-group aggregation (Methods). In Eq. \ref{eq:dynamics_monomer}, monomers disappear due to the formation of a new group from two monomers (first term, Fig. \ref{fig4}a) or due to the growth of an existing group by addition of a new monomer (second term, Fig. \ref{fig4}b). Similarly, in Eq. \ref{eq:dynamics_group}, $c_{\rm g}$ can change due to the formation of a new group from monomers (first term,  Fig. \ref{fig4}a) or from the merging of two existing groups (second term, Fig. \ref{fig4}c). In all cases, we assume that the rate of encounters is proportional to the product of the number densities of the species involved (either monomers or groups) and that any dependence on the effective cross-sectional area of each species is accounted for by the effective rates of aggregation $\alpha_\mathrm{mm}$, $\alpha_\mathrm{mg}$ and $\alpha_\mathrm{gg}$. Without a shared environmental memory, these rates should only depend on the effective diffusion coefficients of the species involved \cite{krapivsky2010kinetic}. The larger the effective diffusion coefficient, the faster is the rate of group formation and growth, leading to larger groups within the same time interval. Nonetheless, our experimental results suggest that the effective diffusion coefficients decrease with $\rho_\mathrm{p}$ (Fig. \ref{fig:MSD}), so that the augmented group formation for intermediate values of $\rho_\mathrm{p}$ must result from the presence of spatial correlations in the environment that increase the chances for particles to meet, i.e. the reuse of transient paths highlighted in Figs. \ref{fig1} and \ref{fig:Extra_Fig1_time_sequence}.

By calculating the effective rate coefficients ($\alpha_\mathrm{mm}$, $\alpha_\mathrm{mg}$ and $\alpha_\mathrm{gg}$) from the simulated data, we can assess the relative importance of monomer--monomer, monomer--group and group--group aggregation on the kinetics of group formation (Fig. \ref{fig4}d). Both $\alpha_\mathrm{mg}$ and $\alpha_\mathrm{gg}$ present a maximum for intermediate values of $\rho_{\rm p}$ when group formation is enhanced, while $\alpha_\mathrm{mm}$ is roughly constant in comparison. In the presence of passive particles, although monomer--monomer aggregation is key for the formation of the initial groups, the kinetics are dominated by groups catalysing their own growth through the addition of new monomers and merging to other existing groups. These aggregation events mediated by the presence of a shared environmental memory are hence behind the enhanced group formation observed at intermediate $\rho_{\rm p}$. Indeed, when the torque is switched off in simulation (Fig.~\ref{fig4}e), the shared memory and stigmergy cannot develop (Fig.~\ref{fig3}), resulting in aggregation rates that decay monotonically with increasing values of $\rho_{\rm p}$, as one would expect when an increased number of obstacles hinders diffusion (Fig. \ref{fig:MSD}).

In summary, our results demonstrate how, in a decentralised system composed of clueless active units with no explicit signalling pathway or information processing capability, a dynamic environment can create the conditions for the emergence of a shared environmental memory that can coordinate and shape the system's collective response. Hence, confinement by crowding becomes a condition sufficient for self-organisation to emerge and to activate the system's coordination capabilities (e.g., by naturally evolving to larger groups) \cite{araujo2023steering}. Similar mechanisms of shared memory which are primarily promoted and reinforced by dynamic environmental factors could also contribute to shaping the collective dynamics of other decentralised systems where individuals can instead actively signal to each other \cite{marsh2008stigmergic}, such as communities of micro-organisms \cite{mukherjee2019bacterial,Austin2020}, social insects \cite{czaczkes2015trail,Reid2011} and robotic swarms \cite{moussaid2009collective,dorigo2020reflections}. The feedback from the environment could then lower the threshold for quorum formation in natural communities \cite{mukherjee2019bacterial} and for reaching consensus in decision-making processes \cite{garnier2007biological}, e.g., by synergistically catalysing any pathway of explicit communication. Finally, we envisage that shared memories promoted by environmental dynamic features could become design factors to implement low-level rules to drive high-level self-organisation in artificial systems, including in the design of antimicrobial surfaces, of crowd management control tools, and of neuromorphic computers and artificial swarm intelligence \cite{araujo2023steering}.

\bmhead{Acknowledgments}

We are grateful to Samantha Rueber and Matthew Blunt for initial training on experimental techniques. CSD and NAMA acknowledge financial support from the Portuguese Foundation for Science and Technology (FCT) under Contracts no. PTDC/FIS-MAC/5689/2020,  EXPL/FIS-MAC/0406/2021, CEECIND/00586/2017, UIDB/00618/2020,  and UIDP/00618/2020. GV (Giorgio) acknowledges sponsorship for this work by the US Office of Naval Research Global (Award No. N62909-18-1-2170). NAMA and GV (Giorgio) acknowledge support from the UCL MAPS Faculty Visiting Fellowship programme.

\bmhead{Author contributions}

Author contributions are defined based on the CRediT (Contributor Roles Taxonomy) and listed alphabetically. Conceptualisation: GV (Giovanni), NAMA, GV (Giorgio). Data Curation: CSD, MT. Formal analysis: CSD, MT, NAMA, GV (Giorgio). Funding acquisition: GV (Giorgio). Investigation: CSD, MT, NAMA. Methodology: CSD, MT, GV (Giovanni), NAMA, GV (Giorgio). Project administration: GV (Giorgio). Resources: NAMA, GV (Giorgio). Software: CSD. Supervision: CSD, NAMA, GV (Giorgio). Validation: CSD, MT, GV (Giorgio). Visualisation: CSD, MT, GV (Giorgio). Writing – original draft: CSD, GV (Giovanni), NAMA, GV (Giorgio). Writing – review and editing: CSD, MT, GV (Giovanni), NAMA, GV (Giorgio).

\bmhead{Competing interests}

The authors have no competing interests.

\bmhead{Data availability}

All data in support of this work is available in the manuscript or the supplementary materials. Further data and materials are available from the corresponding authors upon reasonable request.

\bmhead{Code availability}

The code that supports the findings of this study is available from the corresponding authors upon reasonable request.

\section*{Methods}

\subsection*{Materials}
Glass microscopy slides (Thermo Fisher) were purchased from VWR while glass coverslips were purchased from Thorlabs. The following chemicals were purchased and used as received: 2,6-lutidine ($\ge$ 99$\%$, Sigma-Aldrich), acetone ($\ge$ 99.8$\%$, Sigma-Aldrich), ethanol ($\ge$ 99.8$\%$, Fisher Scientific), sodium hydroxide (NaOH, Fisher Scientific). Deionised (DI) water ($\ge$ 18 M$\Omega.$cm) was collected from a Milli-Q purification system. Aqueous colloidal dispersions (5$\%$ w/v) of silica (${\rm SiO_2}$) colloids for sample preparation (4.77 $\pm$ 0.20 $\mu$m in diameter for Janus particle fabrication and as passive colloids; 7.00 ± 0.15 $\mu$m in diameter as spacers) were purchased from Microparticles GmbH. Carbon rods of length 300 mm and diameter 6.15 mm for coating Janus particles by sputtering were purchased from Agar Scientific and cut to a length of 50 mm before use. Lens tissue for slide cleaning was purchased from Thorlabs. UV cure adhesive (Blufixx) and hydrophobic coating (RainX) for sample preparation were purchased from an online retailer (Amazon).

\subsection*{Slide cleaning protocol}
Before their use for sample preparation, glass slides and coverslips were cleaned by wiping with acetone-soaked lens tissue. RainX (a commercial solution which renders glass surfaces more hydrophobic and aids limiting particle sticking to the glass chamber) was then smeared on both with a cotton bud and gently dried with a nitrogen gun. After 2 min, excess RainX was removed by wiping with acetone-soaked lens tissue. Glass slides for the deposition of colloidal monolayers were instead cleaned by sonication for 10 min in a 2 M NaOH ethanolic solution followed by three cycles of 5 min sonication in DI water. To dry them, the slides were withdrawn from water in the presence of ethanol vapor (Marangoni drying) and, subsequently, blown dry with a nitrogen gun.

\subsection*{Fabrication of Janus particles} 
The Janus particles used in our experiments were fabricated from ${\rm SiO_2}$ colloids of diameter $d = 4.77 \pm 0.20 \, {\rm \mu m}$, which were coated on one side with a thin layer ($\approx60 \, {\rm nm}$) of carbon. We first deposited a monolayer of colloids on a clean glass slide. The monolayer was obtained by evaporating a $40 \, {\rm \mu L}$ droplet containing a $2.5 \% \, {\mathrm w}/{\mathrm v}$ dispersion of the colloids in DI water. The particles were then coated with a 60 nm thick carbon layer using an automatic carbon coater (AGB7367A, Agar Scientific). The thickenss of the carbon layer was confirmed by atomic force measurements (AFM). Post-coating sonication allowed us to dislodge the half-coated particles in DI water from the glass slide to use them for sample preparation.

\subsection*{Sample preparation} 

Samples were prepared in the form of a quasi-two-dimensional glass chamber filled with a colloidal dispersion in a critical mixture of water-2,6-lutidine. Typical colloidal dispersions include Janus particles as well as passive ${\rm SiO_2}$ particles and spacers. For example, to achieve a typical dispersion with packing fractions of $\rho_{\rm a} = 0.5\%$ and $\rho_{\rm p} = 12.5\%$, we mixed stock dispersions of the three types of particles in DI water to achieve an aqueous dispersion containing $0.13 \%$ w/v of Janus particles, $5 \%$ w/v of passive particles and $0.08 \%$ w/v of spacers.
This concentration of spacers was chosen to minimise their number in the field of view, whilst giving enough support to maintain the quasi-two-dimensional chamber's geometry. Samples with other packing fractions ($\rho_{\rm a}$ and $\rho_{\rm p}$) were obtained by linearly scaling these concentrations of Janus particles and passive particles to obtain the right values of $\rho_{\rm a}$ and $\rho_{\rm p}$. Before their use, the colloidal mixtures were centrifuged at 1000 relative centrifugal force (RCF) for 3 minutes leaving a pellet; the supernatant was then removed and replaced with a $28.6 \%$ w/v water-2,6-lutidine solution. This process was repeated three times to remove residual DI water from the initial dispersion. Experiment-ready quasi-two-dimensional sample chambers containing a dispersion of colloids in a critical water-2,6-lutidine solution were prepared by sandwiching ${10 \, \mu {\rm L}}$ of this final dispersion between a clean glass slide and a thin coverslip. The chamber was sealed by applying an ultraviolet-curable adhesive around the edges of the coverslip, which was then exposed to UV light for 30 s on each side. Before data acquisition, the sample was left to equilibrate over a 30 min period.

\subsection*{Optical setup and microscopy} 

All the experiments were performed on an inverted microscope (Leica, DMI4000) equipped with a homemade flow thermostat to maintain the critical suspension at a fixed temperature (T = 30 $^{\circ}$C) below the critical point ($T_{\rm c} \approx 34.1 \, ^{\circ}$C). The sample was illuminated with a green continuous-wave laser ($\lambda = 532 \, {\rm nm}$) at a power density of $2.5 \, {\rm \mu W \, \mu m^{-2}}$ to propel the Janus particles via self-diffusiophoresis due to light absorption at the carbon cap \cite{buttinoni2012active}. Both Janus and passive particles are tracked by digital video microscopy \cite{crocker1996methods} using the image projected by a microscope objective ($\times 20$, ${\rm NA} = 0.5$) on a monochrome complementary metal–oxide–semicondutor (CMOS) camera (Thorlabs, DCC1545M) with an acquisition rate of 10 frames per second. The incoherent illumination for the tracking is provided by a white-light-emitting diode (Thorlabs, MWWHLP1) directly projected onto the sample. A long-pass dichroic mirror (Thorlabs, DMLP605) with a 605-nm cut-on wavelength is used to combine laser and white light before the sample, while laser light is removed from the detection path with a notch filter centred at 532 nm (Semrock, NF01-532U-25).

\subsection*{Path revival function}
To quantify the path reuse by the Janus particles, we define the path revival function $1-C_{\rm aa}(\tau)$, where $C_{\rm aa}(\tau)$ is the cumulative probability that a region crossed by an active particle will be crossed by another particle within a lag time $\tau$. To compute $1-C_{\rm aa}(\tau)$ we define a circular region of diameter $d$ around each active particle at a certain time $t$ and measure how many of those regions have been crossed by the center of another active particle up to lag time $\tau$. If we consider the particles' velocities to be Poisson distributed when a path is chosen, then this function should follow an exponential distribution for persistent particles \cite{enns1984hitting} 
\begin{equation}
1-C_{\rm aa}(\tau)=\exp{\left(-\tau/\tau_{\rho_{\rm p}}\right)},
\end{equation}
where $\tau_{\rho_{\rm p}}$ is the effective path revival lifetime, which we fit from the data. For both experiments and simulations, we assume that the initial positions in the particles' trajectories are uncorrelated (i.e., in the experiments, we only consider trajectories of individual particles before groups form and, in the simulations, the short-range attractive interaction between particles is turned off to prevent group formation).  

\subsection*{Particle-based simulations}
We consider a numerical model where $n_{\rm a}$ active spheres and $n_{\rm p}$ passive spheres of diameter $d$ move inside a two-dimensional box of side $L = 60 d$ with periodic boundary conditions. Both $n_{\rm a}$ and $n_{\rm p}$ are fixed to match the experimental values of $\rho_{\rm a}$ and $\rho_{\rm p}$. As in the experiments, all particles, whether active or passive, have the same size $d$ and mass $m$. 

To map the simulations to the experiments, we consider the same Péclet number defined as,
\begin{equation}    P_\mathrm{e}=\frac{d v}{D_\mathrm{t}},
\end{equation}
where $d=4.77 \, {\rm \mu m}$, $v = 1.9 \, {\rm \mu m \, s^{-1}}$, and $D_\mathrm{t}=0.0249 \, {\rm \mu m^2 \, s^{-1}}$. Both velocity $v$ and diameter $d$ of the active particles were used to convert the reduced units in simulations to SI units. The translational diffusion coefficient $D_\mathrm{t}$ was calculated as
\begin{equation}    D_\mathrm{t}=\frac{k_{\rm B}T}{\gamma'},
\end{equation}
where $k_{\rm B}$ is the Boltzmann constant, $T$ the absolute temperature, and $\gamma'$ is the corrected translational drag coefficient for colloids at distance $s$ from a surface \cite{Leach2009}, given by
\begin{equation}
 \gamma'=\frac{\gamma}{1-(9/16)(d/2s)+(1/8)(d/2s)^{1/3}},
\end{equation}
with 
\begin{equation}
    \gamma=3\pi\mu d,
\end{equation}
where $\mu$ is the fluid viscosity. We assume $2s = d$, $T=303 \, {\rm K}$, and $\mu=2.1\times10^{-3} \, {\rm Pa \, s}$ for the water-lutidine mixture. Similarly, the rotational diffusion coefficient $D_\mathrm{r}$ was calculated as
\begin{equation}    D_\mathrm{r}=\frac{k_{\rm B}T}{\beta'},
\end{equation}
where $\beta'$ is the corrected rotational drag coefficient for colloids at distance $s$ from a surface \cite{Leach2009}, given by
\begin{equation}
 \beta'=\frac{\beta}{1-(1/8)(d/2s)^{1/3}},
\end{equation}
with 
\begin{equation}
    \beta=\pi\mu d^3.
\end{equation}
The trajectories of both active and passive particles were obtained by integrating their equations of motion using a velocity Verlet scheme implemented
in the Large-scale Atomic/Molecular Massively Parallel Simulator
(LAMMPS) \cite{PLIMPTON1995}. Specifically, the particles' translational motion and rotational motion around one single axis (perpendicular to the simulation plane) are respectively described by the following Langevin equations,

\begin{equation}
 m\dot{\vec{v_i}}(t)=-\nabla_{\vec{r_i}} V_i-\frac{m}{\tau_{\gamma}}\vec{v_i}(t)+\sqrt{\frac{2mk_{\rm B}T}{\tau_{\gamma}}}\vec{\xi}_{\rm t}^i(t)+F_{\rm a}\hat{u}_i(t) \label{eq.trans_Langevin_dynamics}
\end{equation}
and 
\begin{equation}
 I\dot{\omega_i}(t)=\Omega_i-\frac{\alpha I}{\tau_{\gamma}}\omega_i(t)+\sqrt{\frac{2\alpha Ik_{\rm B}T}{\tau_{\gamma}}}\xi_{\rm r}^i(t),\label{eq.rot_Langevin_dynamics}
\end{equation}
where $\vec{v_i}$ and $\omega_i$ 
are the translational and angular velocity for particle $i$, $\hat{u}_i=(\cos\theta_i,\sin\theta_i)$, $\omega_i = \dot{\theta}_i$, $F_{\rm a}$ is the strength of the self-propulsion force for the active particles, $\tau_{\rm \gamma}$ is the damping time, $I$ is the particles' inertia of rotation, $V_i$ is the potential due to the interaction with all surrounding particles, and $\Omega_i$ is an effective torque due to the interaction of particle $i$ with the surrounding passive particles. $\vec{\xi_t^i}(t)$ and $\vec{\xi_r^i}(t)$ are stochastic terms taken from a truncated random distribution of zero mean
and unitary standard deviation~\cite{Dunweg1991}. Moreover, $\alpha$ is a model parameter that defines the relationship between the rotational ($D_{\rm r}$) and translational ($D_{\rm t}$) diffusion coefficients as
\begin{equation}
\frac{D_\mathrm{t}}{D_\mathrm{r}}=\alpha \frac{I}{m}.
\end{equation}
where $\alpha$ is adjusted to map the experimental relation between $D_\mathrm{t}$ and $D_\mathrm{r}$.

The motion of the passive particles is only governed by Eq. \ref{eq.trans_Langevin_dynamics}, where we set $F_{\rm a} = 0$. 

The interaction between particles is implemented with a Lennard-Jones potential given by
\begin{equation}
V_i=\sum_jV_{ij}(r_{ij})=\sum_j 4\epsilon_\mathrm{LJ}\left[\left(\frac{\sigma_\mathrm{LJ}}{r_{ij}}\right)^{12}-\left(\frac{\sigma_\mathrm{LJ}}{r_{ij}}\right)^6\right],
\label{eq.LJ}
\end{equation}
where $r_{ij}=\lvert\lvert\vec{r}_i-\vec{r}_j\rvert\rvert$ is the distance between two particles, $\epsilon_\mathrm{LJ}$ the depth of the potential well, and $\sigma_\mathrm{LJ}$ the width of the potential (distance at which the potential is zero). For passive particles, with purely repulsive interactions, we consider a truncated Lennard-Jones potential where the cut-off is set at $r_{\rm cut} = d = 2^{1/6}\sigma_\mathrm{LJ}$ to remove the attractive part. For active particles, we consider an attractive interaction with a cut-off set at $r_{\rm cut} = 5d$. The depth of the potential well $\epsilon_\mathrm{LJ}$ is obtained from experimental data (Fig. \ref{fig:LJ_Cmax}).

Finally, to describe the impact of the passive particles on the rotational degrees of freedom of the active particles, we introduce the effective torque $\Omega_i$ on particle $i$ \cite{Liebchen2019}
\begin{equation}
\Omega_i=-\Omega_0d^2 \hat{v}_i\times\sum_{j=1}^{N_p}\nabla_{\vec{r}}\frac{e^{-\kappa r_{ij}}}{r_{ij}},
\label{eq.torque}
\end{equation}
where $\Omega_0$ sets the strength of the torque, $\hat{v}_i=\vec{v}_i/\lvert\lvert\vec{v}_i\lvert\lvert$ and $\times$ is the cross product. The negative sign indicates that active particle steer away from the surrounding  passive ones; $\kappa=0.25/d$ gives the screening number in agreement with the range of experimental values estimated in \cite{Liebchen2019}. For numerical efficiency, we set a cut-off radius of four particles diameters, where the value of the torque is much lower than the typical thermal noise. The torque used to map the experiments ($\Omega_0 = 72\pm16 \, k_{\rm B} T$) was computed and confirmed from two different experimental measurements (Figs. \ref{fig3}a,c).

\subsection*{Rate equations}
The relevant mechanisms for the dynamics are: (1) the formation of new groups by combining two free active particles (monomers); (2) the growth of a group by the addition of a monomer; 
(3) the pairwise merging of groups; and (4) their fragmentation. In the experiments, we define groups as clusters of size larger than two as dimers are unstable in time. Here, for completeness, we consider all cases. We assume that groups only lose one active particle at a time (fragmentation). 

We define $c_1$ and $c_i$ as the number densities of free active particles (monomers) and groups of size $i>1$, respectively. The following rate equations then gives the time evolution of $c_1$,
\begin{equation}
\dot{c}_1=-\alpha_\mathrm{mm}c^2_1-\alpha_\mathrm{mg}c_1\sum_{j>1}c_j +
\sum_{j>1}f_jc_j+f_2c_2,
\end{equation}
where the first term accounts for the formation of new groups, the second one for the growth of an existing group, the third for fragmentation, and the fourth term for the additional free active particle obtained from the fragmentation of groups of size two. If the distance between groups is larger than the persistence length of the free active particles, the main mechanism of mass transport is diffusion and, in the absence of spatial correlations, the rates $\alpha_\mathrm{mm}$ and $\alpha_\mathrm{mg}$ should only depend on the size of the active particles and on their effective diffusion coefficients~\cite{krapivsky2010kinetic}. For simplicity, we also consider that $\alpha_\mathrm{mg}$ does not depend on the group size. 

Similarly, for groups of size two,

\begin{eqnarray}
\dot{c}_2=\frac{\alpha_\mathrm{mm}}{2}c^2_1-\alpha_\mathrm{mg}c_1c_2-\alpha_\mathrm{gg}c_2\sum_{j>1}c_j-f_2c_2+f_3c_3,\ \
\end{eqnarray}
and for groups of size $k$,

\begin{eqnarray}
\dot{c}_k=\alpha_\mathrm{mg}c_1c_{k-1}-\alpha_\mathrm{mg}c_1c_k+\frac{1}{2}\alpha_\mathrm{gg}\sum_{i+j=k}c_ic_j-\alpha_\mathrm{gg}c_k\sum_{j>1}c_j+f_{k+1}c_{k+1}-f_kc_k. \ \
\end{eqnarray}

If we now define the number density of groups $c_\mathrm{g}=\sum_{j>1} c_j$ and the total fragmentation rate $f=\sum_{j>1}f_jc_j$, we obtain,

\begin{equation}
    \dot{c}_1=-\alpha_\mathrm{mm}c_1^2-\alpha_\mathrm{mg}c_1c_{\rm g}+f+f_2c_2 \\ ,
\end{equation}

\noindent and,

\begin{equation}
    \dot{c}_{\rm g}=\frac{\alpha_\mathrm{mm}}{2}c_1^2-\frac{\alpha_\mathrm{gg}}{2}c_{\rm g}^2-f_2c_2 \\ .
\end{equation}

In simulation, we observe that the total fragmentation rate is constant for a large range of $\rho_\mathrm{p}$ up to the intermediate values where the largest groups are observed and then drops fast at higher values (Fig. \ref{fig:fragmentation}). Thus, aggregation rather than fragmentation is the leading factor in the non-monotonic dynamics of group formation observed in Fig. \ref{fig2}. If we neglect fragmentation, we obtain Eqs. \ref{eq:dynamics_monomer} and \ref{eq:dynamics_group}, respectively.



\bibliography{sn-article}

\clearpage
\setcounter{page}{0}
\pagenumbering{arabic}
\setcounter{page}{1}

\section*{Supplementary Information}

\renewcommand{\thefigure}{S\arabic{figure}}
\setcounter{figure}{0}

\subsection*{Supplementary Figures}

\begin{figure}[htb!]
    \centering
    \includegraphics[width=0.4\textwidth]{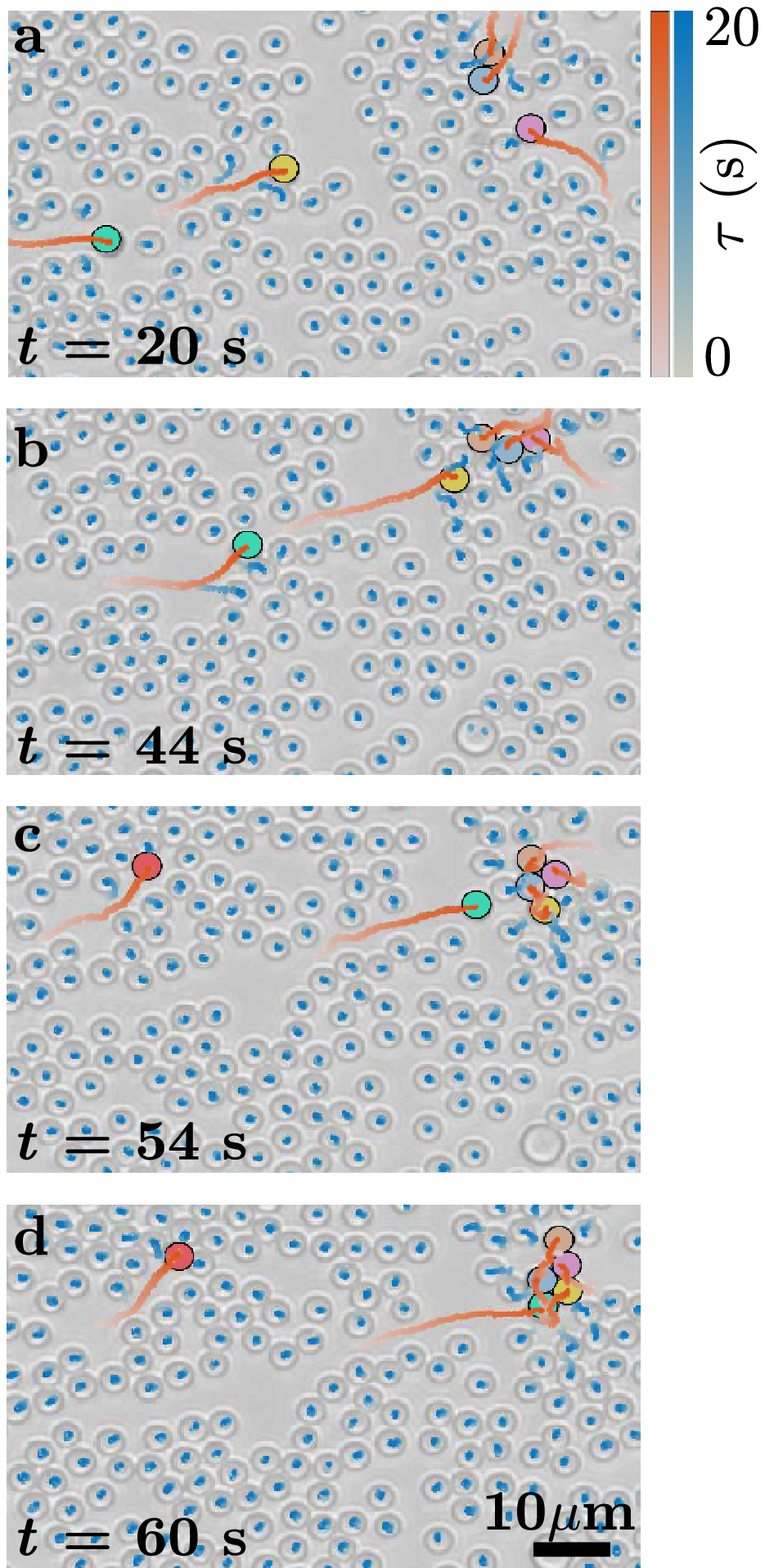}
    \caption{\textbf{Additional example of group formation of active particles via environmental memory.} 
    (\textbf{a-d}) Time sequence showing (\textbf{a-b}) a light-activated Janus particle (yellow) joining a newly formed group of three active particles (orange, purple and blue) after having burrowed through a crowded environment of ${\rm SiO_2}$ passive particles (densities {$\rho_{\rm a} = 1.1\%$} and $\rho_{\rm p} = 37.5\%$ as in Fig. \ref{fig1}). (\textbf{c-d}) A nearby Janus particle (cyan) reuses this path before joining the same cluster. In each image, $20$ s-long trajectories are shown for both active (red) and passive (blue) particles.} 
    \label{fig:Extra_Fig1_time_sequence}
\end{figure}

\begin{figure}[htb!]
     \centering
     \includegraphics[width=0.6\textwidth]{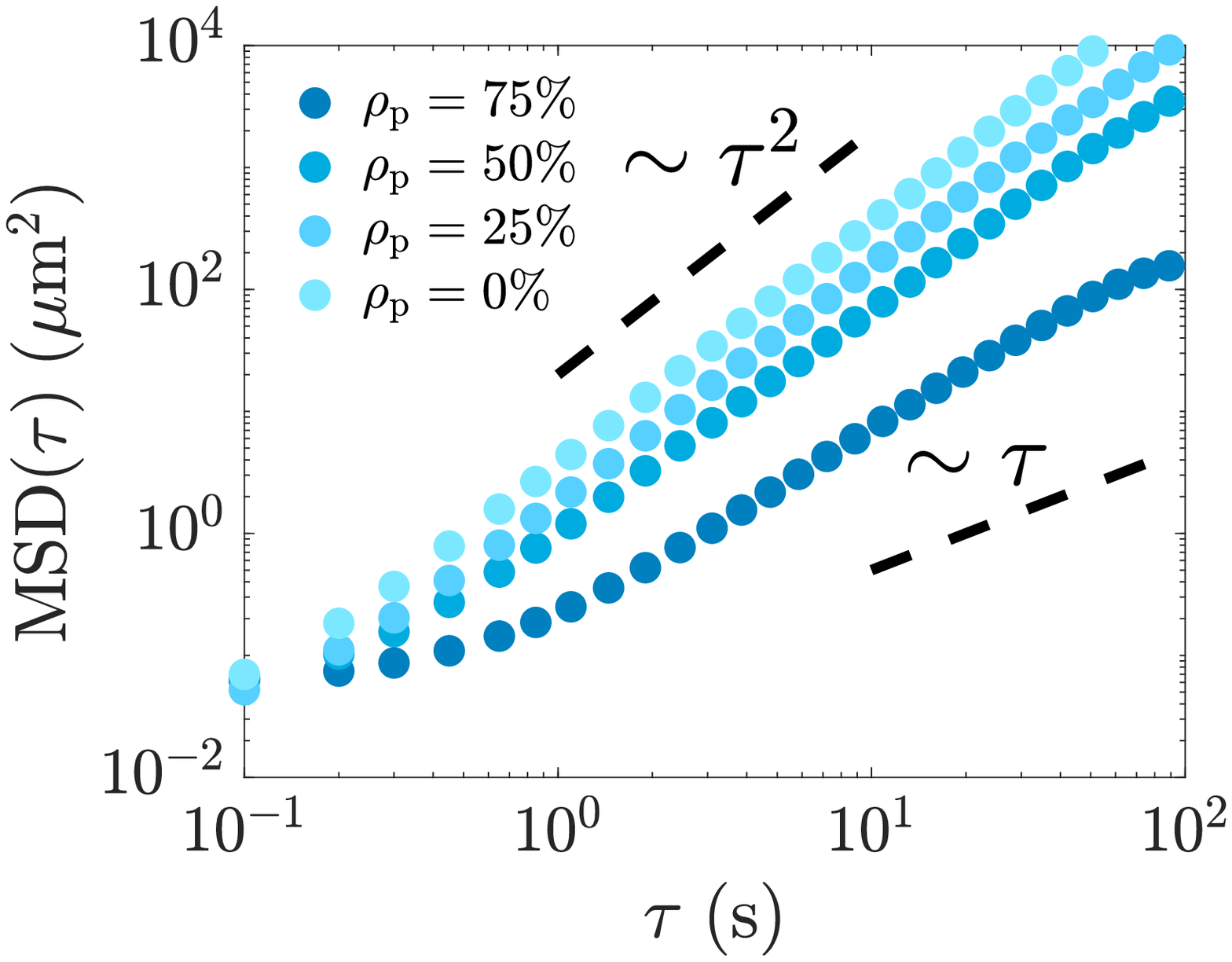}
     \caption{\textbf{Mean square displacements of Janus colloids for different $\rho_{\rm p}$.} Average mean square displacements (MSDs) of active colloids ($\rho_{\rm a} = 1.1\%$ as in Figs. \ref{fig1} and \ref{fig:Extra_Fig1_time_sequence}) self-propelling through different densities $\rho_{\rm p}$ of passive colloids. The MSDs show that the area explored by the active colloids decreases when $\rho_{\rm p}$ increases due to the additional resistance imposed by the passive colloids on the active particles' motion. The two dashed lines show persistent ($\propto \tau^2$) and diffusive ($\propto \tau$) behaviour, respectively, for reference. 
     Each experimental MSD curve was obtained as an ensemble average over the trajectories of at least 143 Janus particles.} 
     \label{fig:MSD}
\end{figure}

\begin{figure}[htb!]
     \centering     \includegraphics[width=0.9\textwidth]{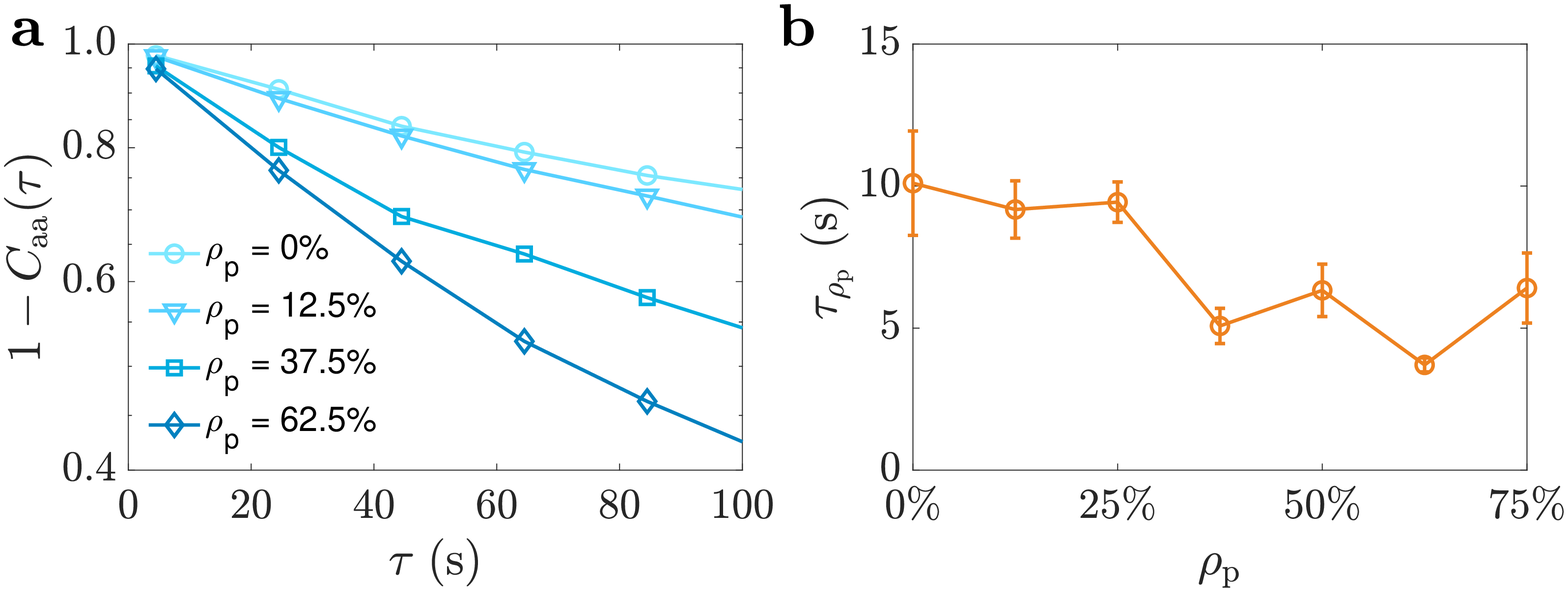}
     \caption{\textbf{Path revival function and lifetime at different $\rho_{\rm p}$.} (\textbf{a}) The path revival function $1-C_{\rm aa}(\tau)$ calculated for $\rho_{\rm a} = 1.1\%$ at different densities $\rho_{\rm p}$ of passive colloids shows the probability that a region visited by an active colloid will be revisited by some other active particle within a given lag time $\tau$ before a group is formed. The faster the decay of this function, the longer a previously formed path survives in time as its reuse stabilises it for longer before it closes due to thermal fluctuations. Each experimental revival function was obtained as an ensemble average over the trajectories of three videos at $\rho_{\rm a} = 1.1\%$ per each $\rho_{\rm p}$ value. (\textbf{b}) Fitting the previous trends to an exponential $1-C_{\rm aa}(\tau) = \exp{\left(-\tau/\tau_{\rho_{\rm p}}\right)}$, we can calculate an effective path revival lifetime $\tau_{\rho_{\rm p}}$ as a function of $\rho_{\rm p}$, which indeed initially decreases with $\rho_{\rm p}$ despite the particles' motion becoming hampered by higher levels of crowding (Fig.~\ref{fig:MSD}). The error bars are an estimate of the uncertainty of the fitting.} 
     \label{fig:lifetime}
\end{figure}

\begin{figure}[htb!]
    \centering
    \includegraphics[width=0.8\textwidth]{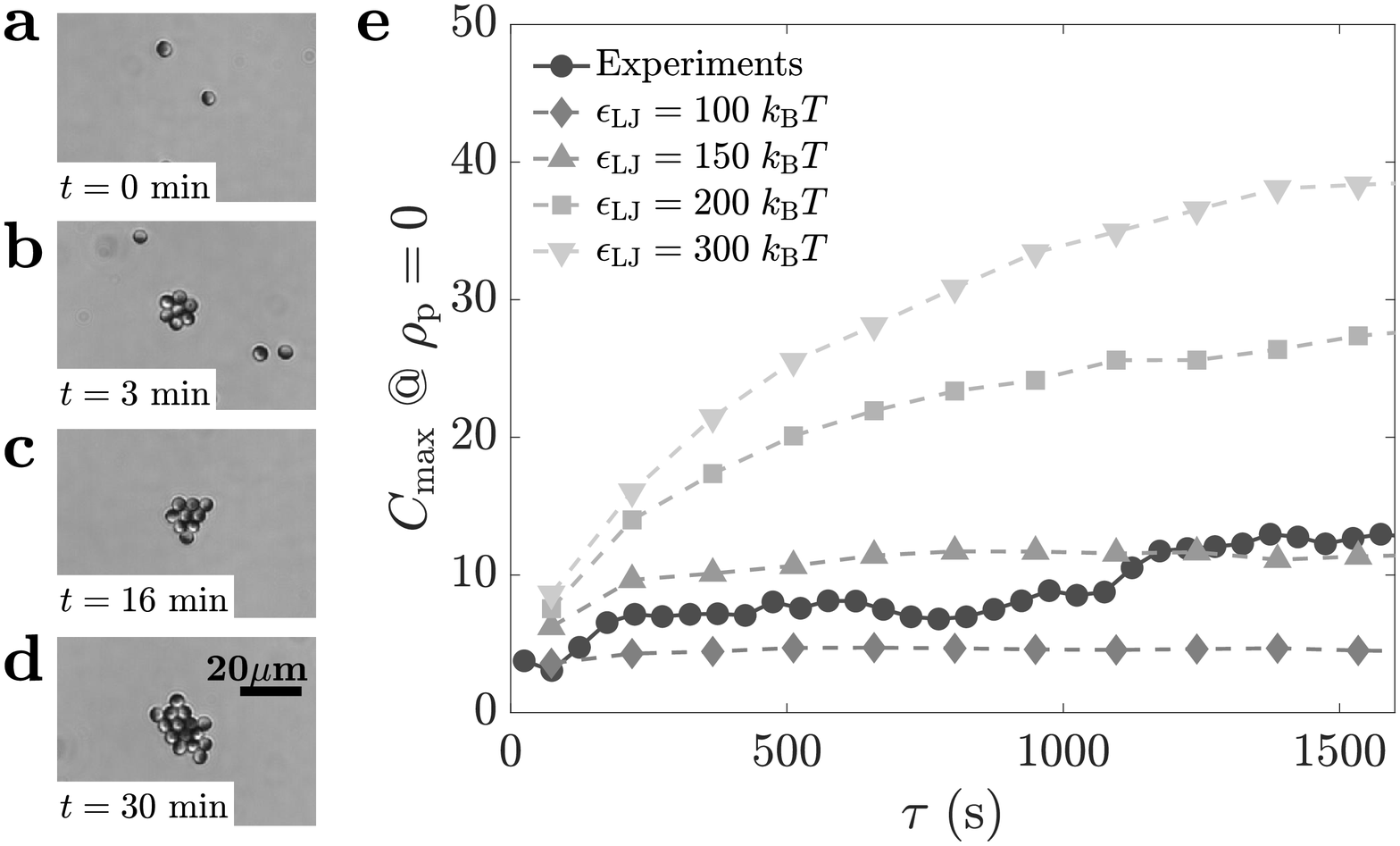}
    \caption{ \textbf{Estimation of the Lennard-Jones potential's depth, $\epsilon_{\rm LJ}$.} ({\bf a-d}) Snapshots of a time sequence showing the formation of a group in a homogeneous environment ($\rho_{\rm p} = 0\%$). ({\bf e}) The change of this group's size $C_{\rm max}$ in time allow us to estimate the depth $\epsilon_{\rm LJ}$ of the Lennard-Jones potential between active colloids by matching the simulations to this experimental trend. The experimental results are better reproduced for $\epsilon_{\rm LJ} = 150 \, k_{\rm B} T$.} 
    \label{fig:LJ_Cmax}
\end{figure}

\begin{figure}[htb!]
    \centering    \includegraphics[width=0.6\textwidth]{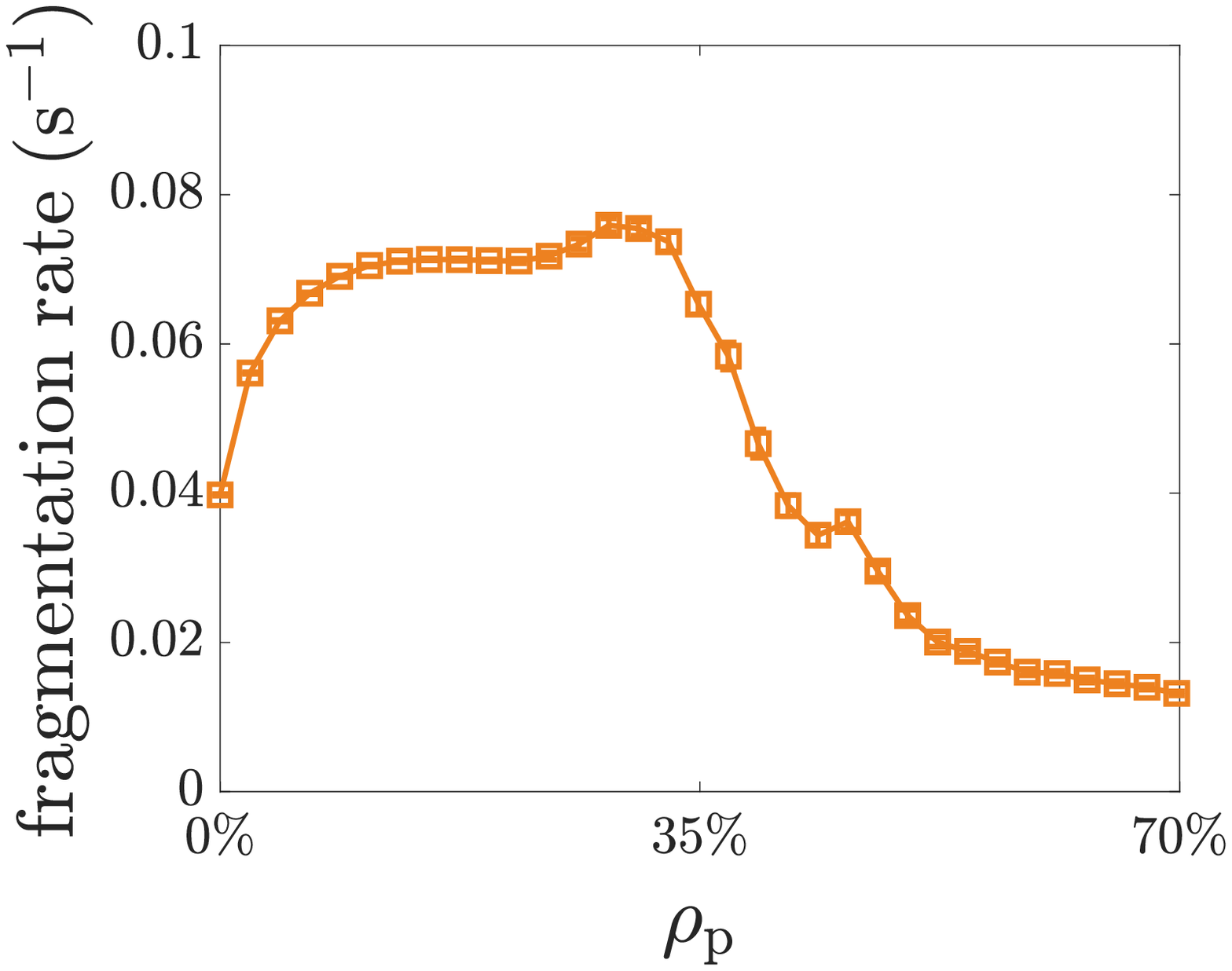}
    \caption{ \textbf{Fragmentation rate as a function of $\rho_{\rm p}$.} Total fragmentation rate accounting for all possible fragmentation processes (calculated as the number of fragmentation events per unit of time) for the calculations shown in Fig. \ref{fig4}. This rate is approximately constant for low and intermediate values of $\rho_{\rm p}$, i.e. where the largest groups are observed in our experiments, and drops fast at higher values. This supports aggregation, rather than fragmentation, as the leading factor in the dynamics of group formation observed in Fig. \ref{fig2}.} 
    \label{fig:fragmentation}
\end{figure}

\end{document}